\documentclass{aastex62}

\usepackage{txfonts}

\shorttitle{GW counterpart search by CALET}
\shortauthors{Adriani et al.}

\begin{document}

\title{Search for GeV Gamma-ray Counterparts of Gravitational Wave Events by {\sl CALET}}

\author{O.~Adriani}
\affil{Department of Physics, University of Florence, Via Sansone, 1 - 50019 Sesto, Fiorentino, Italy}
\affil{INFN Sezione di Florence, Via Sansone, 1 - 50019 Sesto, Fiorentino, Italy}
\author{Y.~Akaike}
\affil{Department of Physics, University of Maryland, Baltimore County, 1000 Hilltop Circle, Baltimore, MD 21250, USA}
\affil{Astroparticle Physics Laboratory, NASA/GSFC, Greenbelt, MD 20771, USA}
\author{K.~Asano}
\affil{Institute for Cosmic Ray Research, The University of Tokyo, 5-1-5 Kashiwa-no-Ha, Kashiwa, Chiba 277-8582, Japan}
\author{Y.~Asaoka}
\affil{Research Institute for Science and Engineering, Waseda University, 3-4-1 Okubo, Shinjuku, Tokyo 169-8555, Japan}
\affil{JEM Utilization Center, Human Spaceflight Technology Directorate, Japan Aerospace Exploration Agency, 2-1-1 Sengen, Tsukuba, Ibaraki 305-8505, Japan}
\author{M.G.~Bagliesi}
\affil{Department of Physical Sciences, Earth and Environment, University of Siena, via Roma 56, 53100 Siena, Italy}
\affil{INFN Sezione di Pisa, Polo Fibonacci, Largo B. Pontecorvo, 3 - 56127 Pisa, Italy}
\author{E.~Berti}
\affil{Department of Physics, University of Florence, Via Sansone, 1 - 50019 Sesto, Fiorentino, Italy}
\affil{INFN Sezione di Florence, Via Sansone, 1 - 50019 Sesto, Fiorentino, Italy}
\author{G.~Bigongiari}
\affil{Department of Physical Sciences, Earth and Environment, University of Siena, via Roma 56, 53100 Siena, Italy}
\affil{INFN Sezione di Pisa, Polo Fibonacci, Largo B. Pontecorvo, 3 - 56127 Pisa, Italy}
\author{W.R.~Binns}
\affil{Department of Physics, Washington University, One Brookings Drive, St. Louis, MO 63130-4899, USA}
\author{S.~Bonechi}
\affil{Department of Physical Sciences, Earth and Environment, University of Siena, via Roma 56, 53100 Siena, Italy}
\affil{INFN Sezione di Pisa, Polo Fibonacci, Largo B. Pontecorvo, 3 - 56127 Pisa, Italy}
\author{M.~Bongi}
\affil{Department of Physics, University of Florence, Via Sansone, 1 - 50019 Sesto, Fiorentino, Italy}
\affil{INFN Sezione di Florence, Via Sansone, 1 - 50019 Sesto, Fiorentino, Italy}
\author{P.~Brogi}
\affil{Department of Physical Sciences, Earth and Environment, University of Siena, via Roma 56, 53100 Siena, Italy}
\affil{INFN Sezione di Pisa, Polo Fibonacci, Largo B. Pontecorvo, 3 - 56127 Pisa, Italy}
\author{J.H.~Buckley}
\affil{Department of Physics, Washington University, One Brookings Drive, St. Louis, MO 63130-4899, USA}
\author{N.~Cannady}
\affil{Department of Physics and Astronomy, Louisiana State University, 202 Nicholson Hall, Baton Rouge, LA 70803, USA}
\author{G.~Castellini}
\affil{Institute of Applied Physics (IFAC),  National Research Council (CNR), Via Madonna del Piano, 10, 50019 Sesto, Fiorentino, Italy}
\author{C.~Checchia}
\affil{Department of Physics and Astronomy, University of Padova, Via Marzolo, 8, 35131 Padova, Italy}
\affil{INFN Sezione di Padova, Via Marzolo, 8, 35131 Padova, Italy} 
\author{M.L.~Cherry}
\affil{Department of Physics and Astronomy, Louisiana State University, 202 Nicholson Hall, Baton Rouge, LA 70803, USA}
\author{G.~Collazuol}
\affil{Department of Physics and Astronomy, University of Padova, Via Marzolo, 8, 35131 Padova, Italy}
\affil{INFN Sezione di Padova, Via Marzolo, 8, 35131 Padova, Italy} 
\author{V.~Di~Felice}
\affil{University of Rome ``Tor Vergata'', Via della Ricerca Scientifica 1, 00133 Rome, Italy}
\affil{INFN Sezione di Rome ``Tor Vergata'', Via della Ricerca Scientifica 1, 00133 Rome, Italy}
\author{K.~Ebisawa}
\affil{Institute of Space and Astronautical Science, Japan Aerospace Exploration Agency, 3-1-1 Yoshinodai, Chuo, Sagamihara, Kanagawa 252-5210, Japan}
\author{H.~Fuke}
\affil{Institute of Space and Astronautical Science, Japan Aerospace Exploration Agency, 3-1-1 Yoshinodai, Chuo, Sagamihara, Kanagawa 252-5210, Japan}
\author{T.G.~Guzik}
\affil{Department of Physics and Astronomy, Louisiana State University, 202 Nicholson Hall, Baton Rouge, LA 70803, USA}
\author{T.~Hams}
\affil{Department of Physics, University of Maryland, Baltimore County, 1000 Hilltop Circle, Baltimore, MD 21250, USA}
\affil{CRESST and Astroparticle Physics Laboratory NASA/GSFC, Greenbelt, MD 20771, USA}
\author{M.~Hareyama}
\affil{St. Marianna University School of Medicine, 2-16-1, Sugao, Miyamae-ku, Kawasaki, Kanagawa 216-8511, Japan}
\author{N.~Hasebe}
\affil{Research Institute for Science and Engineering, Waseda University, 3-4-1 Okubo, Shinjuku, Tokyo 169-8555, Japan}
\author{K.~Hibino}
\affil{Kanagawa University, 3-27-1 Rokkakubashi, Kanagawa, Yokohama, Kanagawa 221-8686, Japan}
\author{M.~Ichimura}
\affil{Faculty of Science and Technology, Graduate School of Science and Technology, Hirosaki University, 3, Bunkyo, Hirosaki, Aomori 036-8561, Japan}
\author{K.~Ioka}
\affil{Yukawa Institute for Theoretical Physics, Kyoto University, Kitashirakawa Oiwakecho, Sakyo, Kyoto 606-8502, Japan}
\author{W.~Ishizaki}
\affil{Institute for Cosmic Ray Research, The University of Tokyo, 5-1-5 Kashiwa-no-Ha, Kashiwa, Chiba 277-8582, Japan}
\author{M.H.~Israel}
\affil{Department of Physics, Washington University, One Brookings Drive, St. Louis, MO 63130-4899, USA}
\author{K.~Kasahara}
\affil{Research Institute for Science and Engineering, Waseda University, 3-4-1 Okubo, Shinjuku, Tokyo 169-8555, Japan}
\author{J.~Kataoka}
\affil{Research Institute for Science and Engineering, Waseda University, 3-4-1 Okubo, Shinjuku, Tokyo 169-8555, Japan}
\author{R.~Kataoka}
\affil{National Institute of Polar Research, 10-3, Midori-cho, Tachikawa, Tokyo 190-8518, Japan}
\author{Y.~Katayose}
\affil{Faculty of Engineering, Division of Intelligent Systems Engineering, Yokohama National University, 79-5 Tokiwadai, Hodogaya, Yokohama 240-8501, Japan}
\author{C.~Kato}
\affil{Faculty of Science, Shinshu University, 3-1-1 Asahi, Matsumoto, Nagano 390-8621, Japan}
\author{N.~Kawanaka}
\affil{Hakubi Center, Kyoto University, Yoshida Honmachi, Sakyo-ku, Kyoto, 606-8501, Japan}
\affil{Department of Astronomy, Graduate School of Science, Kyoto University, Kitashirakawa Oiwake-cho, Sakyo-ku, Kyoto, 606-8502, Japan}
\author{Y.~Kawakubo}
\affil{College of Science and Engineering, Department of Physics and Mathematics, Aoyama Gakuin University,  5-10-1 Fuchinobe, Chuo, Sagamihara, Kanagawa 252-5258, Japan}
\author{H.S.~Krawczynski}
\affil{Department of Physics, Washington University, One Brookings Drive, St. Louis, MO 63130-4899, USA}
\author{J.F.~Krizmanic}
\affil{CRESST and Astroparticle Physics Laboratory NASA/GSFC, Greenbelt, MD 20771, USA}
\affil{Department of Physics, University of Maryland, Baltimore County, 1000 Hilltop Circle, Baltimore, MD 21250, USA}
\author{K.~Kohri} 
\affil{Institute of Particle and Nuclear Studies, High Energy Accelerator Research Organization, 1-1 Oho, Tsukuba, Ibaraki, 305-0801, Japan} 
\author{T.~Lomtadze}
\affil{INFN Sezione di Pisa, Polo Fibonacci, Largo B. Pontecorvo, 3 - 56127 Pisa, Italy}
\author{P.~Maestro}
\affil{Department of Physical Sciences, Earth and Environment, University of Siena, via Roma 56, 53100 Siena, Italy}
\affil{INFN Sezione di Pisa, Polo Fibonacci, Largo B. Pontecorvo, 3 - 56127 Pisa, Italy}
\author{P.S.~Marrocchesi}
\affil{Department of Physical Sciences, Earth and Environment, University of Siena, via Roma 56, 53100 Siena, Italy}
\affil{INFN Sezione di Pisa, Polo Fibonacci, Largo B. Pontecorvo, 3 - 56127 Pisa, Italy}
\author{A.M.~Messineo}
\affil{University of Pisa, Polo Fibonacci, Largo B. Pontecorvo, 3 - 56127 Pisa, Italy}
\affil{INFN Sezione di Pisa, Polo Fibonacci, Largo B. Pontecorvo, 3 - 56127 Pisa, Italy}
\author{J.W.~Mitchell}
\affil{Astroparticle Physics Laboratory, NASA/GSFC, Greenbelt, MD 20771, USA}
\author{S.~Miyake}
\affil{Department of Electrical and Electronic Systems Engineering, National Institute of Technology, Ibaraki College, 866 Nakane, Hitachinaka, Ibaraki 312-8508 Japan}
\author{A.A.~Moiseev}
\affil{Department of Astronomy, University of Maryland, College Park, Maryland 20742, USA }
\affil{CRESST and Astroparticle Physics Laboratory NASA/GSFC, Greenbelt, MD 20771, USA}
\author{K.~Mori}
\affil{Research Institute for Science and Engineering, Waseda University, 3-4-1 Okubo, Shinjuku, Tokyo 169-8555, Japan}
\affil{Institute of Space and Astronautical Science, Japan Aerospace Exploration Agency, 3-1-1 Yoshinodai, Chuo, Sagamihara, Kanagawa 252-5210, Japan}
\author{M.~Mori}
\affil{Department of Physical Sciences, College of Science and Engineering, Ritsumeikan University, Shiga 525-8577, Japan}
\author{N.~Mori}
\affil{INFN Sezione di Florence, Via Sansone, 1 - 50019 Sesto, Fiorentino, Italy}
\author{H.M.~Motz}
\affil{International Center for Science and Engineering Programs, Waseda University, 3-4-1 Okubo, Shinjuku, Tokyo 169-8555, Japan}
\author{K.~Munakata}
\affil{Faculty of Science, Shinshu University, 3-1-1 Asahi, Matsumoto, Nagano 390-8621, Japan}
\author{H.~Murakami}
\affil{Research Institute for Science and Engineering, Waseda University, 3-4-1 Okubo, Shinjuku, Tokyo 169-8555, Japan}
\author{S.~Nakahira}
\affil{RIKEN, 2-1 Hirosawa, Wako, Saitama 351-0198, Japan}
\author{J.~Nishimura}
\affil{Institute of Space and Astronautical Science, Japan Aerospace Exploration Agency, 3-1-1 Yoshinodai, Chuo, Sagamihara, Kanagawa 252-5210, Japan}
\author{G.A.~de~Nolfo}
\affil{Heliospheric Physics Laboratory, NASA/GSFC, Greenbelt, MD 20771, USA}
\author{S.~Okuno}
\affil{Kanagawa University, 3-27-1 Rokkakubashi, Kanagawa, Yokohama, Kanagawa 221-8686, Japan}
\author{J.F.~Ormes}
\affil{Department of Physics and Astronomy, University of Denver, Physics Building, Room 211, 2112 East Wesley Ave., Denver, CO 80208-6900, USA}
\author{S.~Ozawa}
\affil{Research Institute for Science and Engineering, Waseda University, 3-4-1 Okubo, Shinjuku, Tokyo 169-8555, Japan}
\author{L.~Pacini}
\affil{Department of Physics, University of Florence, Via Sansone, 1 - 50019 Sesto, Fiorentino, Italy}
\affil{Institute of Applied Physics (IFAC),  National Research Council (CNR), Via Madonna del Piano, 10, 50019 Sesto, Fiorentino, Italy}
\affil{INFN Sezione di Florence, Via Sansone, 1 - 50019 Sesto, Fiorentino, Italy}
\author{F.~Palma}
\affil{University of Rome ``Tor Vergata'', Via della Ricerca Scientifica 1, 00133 Rome, Italy}
\affil{INFN Sezione di Rome ``Tor Vergata'', Via della Ricerca Scientifica 1, 00133 Rome, Italy}
\author{P.~Papini}
\affil{INFN Sezione di Florence, Via Sansone, 1 - 50019 Sesto, Fiorentino, Italy}
\author{A.V.~Penacchioni}
\affil{Department of Physical Sciences, Earth and Environment, University of Siena, via Roma 56, 53100 Siena, Italy}
\affil{ASI Science Data Center (ASDC), Via del Politecnico snc, 00133 Rome, Italy}
\author{B.F.~Rauch}
\affil{Department of Physics, Washington University, One Brookings Drive, St. Louis, MO 63130-4899, USA}
\author{S.B.~Ricciarini}
\affil{Institute of Applied Physics (IFAC),  National Research Council (CNR), Via Madonna del Piano, 10, 50019 Sesto, Fiorentino, Italy}
\affil{INFN Sezione di Florence, Via Sansone, 1 - 50019 Sesto, Fiorentino, Italy}
\author{K.~Sakai}
\affil{CRESST and Astroparticle Physics Laboratory NASA/GSFC, Greenbelt, MD 20771, USA}
\affil{Department of Physics, University of Maryland, Baltimore County, 1000 Hilltop Circle, Baltimore, MD 21250, USA}
\author{T.~Sakamoto}
\affil{College of Science and Engineering, Department of Physics and Mathematics, Aoyama Gakuin University,  5-10-1 Fuchinobe, Chuo, Sagamihara, Kanagawa 252-5258, Japan}
\author{M.~Sasaki}
\affil{CRESST and Astroparticle Physics Laboratory NASA/GSFC, Greenbelt, MD 20771, USA}
\affil{Department of Astronomy, University of Maryland, College Park, Maryland 20742, USA }
\author{Y.~Shimizu}
\affil{Kanagawa University, 3-27-1 Rokkakubashi, Kanagawa, Yokohama, Kanagawa 221-8686, Japan}
\author{A.~Shiomi}
\affil{College of Industrial Technology, Nihon University, 1-2-1 Izumi, Narashino, Chiba 275-8575, Japan}
\author{R.~Sparvoli}
\affil{University of Rome ``Tor Vergata'', Via della Ricerca Scientifica 1, 00133 Rome, Italy}
\affil{INFN Sezione di Rome ``Tor Vergata'', Via della Ricerca Scientifica 1, 00133 Rome, Italy}
\author{P.~Spillantini}
\affil{Department of Physics, University of Florence, Via Sansone, 1 - 50019 Sesto, Fiorentino, Italy}
\author{F.~Stolzi}
\affil{Department of Physical Sciences, Earth and Environment, University of Siena, via Roma 56, 53100 Siena, Italy}
\affil{INFN Sezione di Pisa, Polo Fibonacci, Largo B. Pontecorvo, 3 - 56127 Pisa, Italy}
\author{J.E.~Suh}
\affil{Department of Physical Sciences, Earth and Environment, University of Siena, via Roma 56, 53100 Siena, Italy}
\affil{INFN Sezione di Pisa, Polo Fibonacci, Largo B. Pontecorvo, 3 - 56127 Pisa, Italy}
\author{A.~Sulaj}
\affil{Department of Physical Sciences, Earth and Environment, University of Siena, via Roma 56, 53100 Siena, Italy}
\affil{INFN Sezione di Pisa, Polo Fibonacci, Largo B. Pontecorvo, 3 - 56127 Pisa, Italy}
\author{I.~Takahashi}
\affil{Kavli Institute for the Physics and Mathematics of the Universe, The University of Tokyo, 5-1-5 Kashiwanoha, Kashiwa, 277-8583, Japan}
\author{M.~Takayanagi}
\affil{Institute of Space and Astronautical Science, Japan Aerospace Exploration Agency, 3-1-1 Yoshinodai, Chuo, Sagamihara, Kanagawa 252-5210, Japan}
\author{M.~Takita}
\affil{Institute for Cosmic Ray Research, The University of Tokyo, 5-1-5 Kashiwa-no-Ha, Kashiwa, Chiba 277-8582, Japan}
\author{T.~Tamura}
\affil{Kanagawa University, 3-27-1 Rokkakubashi, Kanagawa, Yokohama, Kanagawa 221-8686, Japan}
\author{N.~Tateyama}
\affil{Kanagawa University, 3-27-1 Rokkakubashi, Kanagawa, Yokohama, Kanagawa 221-8686, Japan}
\author{T.~Terasawa}
\affil{RIKEN, 2-1 Hirosawa, Wako, Saitama 351-0198, Japan}
\author{H.~Tomida}
\affil{Institute of Space and Astronautical Science, Japan Aerospace Exploration Agency, 3-1-1 Yoshinodai, Chuo, Sagamihara, Kanagawa 252-5210, Japan}
\author{S.~Torii}
\affil{Research Institute for Science and Engineering, Waseda University, 3-4-1 Okubo, Shinjuku, Tokyo 169-8555, Japan}
\affil{JEM Utilization Center, Human Spaceflight Technology Directorate, Japan Aerospace Exploration Agency, 2-1-1 Sengen, Tsukuba, Ibaraki 305-8505, Japan}
\affil{School of Advanced Science and Engineering, Waseda University, 3-4-1 Okubo, Shinjuku, Tokyo 169-8555, Japan}
\author{Y.~Tsunesada}
\affil{Division of Mathematics and Physics, Graduate School of Science, Osaka City University, 3-3-138 Sugimoto, Sumiyoshi, Osaka 558-8585, Japan}
\author{Y.~Uchihori}
\affil{National Institutes for Quantum and Radiation Science and Technology, 4-9-1 Anagawa, Inage, Chiba 263-8555, JAPAN}
\author{S.~Ueno}
\affil{Institute of Space and Astronautical Science, Japan Aerospace Exploration Agency, 3-1-1 Yoshinodai, Chuo, Sagamihara, Kanagawa 252-5210, Japan}
\author{E.~Vannuccini}
\affil{INFN Sezione di Florence, Via Sansone, 1 - 50019 Sesto, Fiorentino, Italy}
\author{J.P.~Wefel}
\affil{Department of Physics and Astronomy, Louisiana State University, 202 Nicholson Hall, Baton Rouge, LA 70803, USA}
\author{K.~Yamaoka}
\affil{Nagoya University, Furo, Chikusa, Nagoya 464-8601, Japan}
\author{S.~Yanagita}
\affil{College of Science, Ibaraki University, 2-1-1 Bunkyo, Mito, Ibaraki 310-8512, Japan}
\author{A.~Yoshida}
\affil{College of Science and Engineering, Department of Physics and Mathematics, Aoyama Gakuin University,  5-10-1 Fuchinobe, Chuo, Sagamihara, Kanagawa 252-5258, Japan}
\author{K.~Yoshida}
\affil{Department of Electronic Information Systems, Shibaura Institute of Technology, 307 Fukasaku, Minuma, Saitama 337-8570, Japan}

\collaboration{(CALET Collaboration)}
\noaffiliation

\correspondingauthor{Masaki Mori and Yoichi Asaoka}
\email{morim@fc.ritsumei.ac.jp; yoichi.asaoka@aoni.waseda.jp}


\begin{abstract}

We present results on searches for gamma-ray counterparts of the LIGO/Virgo 
gravitational-wave events using CALorimetric Electron Telescope ({\sl CALET}) observations. 
The main instrument of {\sl CALET}, CALorimeter (CAL), observes gamma-rays 
from $\sim1$ GeV up to 10 TeV with a field of view of nearly 2 sr. 
In addition, the {\sl CALET} gamma-ray burst monitor (CGBM) views $\sim$3 sr and
$\sim2\pi$ sr of the sky in the 7 keV -- 1 MeV and the 40 keV -- 20 MeV bands, respectively, by using two
different crystal scintillators. The {\sl CALET} observations on the International Space Station
started in October 2015, and here we report analyses of events 
associated with the following gravitational wave events: GW151226, 
GW170104, GW170608, GW170814 and GW170817. 
Although only upper limits on gamma-ray emission are obtained,
they correspond to a luminosity of $10^{49}\sim10^{53}$ erg s$^{-1}$ in the GeV energy band 
depending on the distance and the assumed time duration of each event, which is approximately the order
of luminosity of typical short gamma-ray bursts. 
This implies there will be a favorable opportunity to detect high-energy gamma-ray emission
in further observations if additional gravitational wave events with favorable geometry 
will occur within our field-of-view.
We also show the sensitivity of {\sl CALET} for gamma-ray transient events which is the order
of $10^{-7}$~erg\,cm$^{-2}$\,s$^{-1}$ for an observation of 100~s duration.
\end{abstract}


\section{Introduction}\label{sec:Intro}

The discovery of gravitational-wave (GW) events using laser interferometers
by the LIGO and Virgo Scientific Collaborations \citep{Abb16a} 
was an epoch-making development following
the prediction of the existence of gravitational waves by \citet{Ein16} a hundred years earlier.
GW events are thought to be produced in the last stage of merging compact binaries,
and electromagnetic counterparts of these events have been
extensively discussed by many authors. Merging
neutron star -- neutron star (NS-NS) binaries and neutron star -- black hole (NS-BH)
binaries are thought to emit significant amount of electromagnetic radiation
(e.g. \citet{Phi09,Ros15,Fer16}),
while it is often assumed that gravitational-wave events 
resulting from the merger of stellar-mass 
black holes are unlikely to produce electromagnetic counterparts
(e.g., \citet{deM17}).
%
 
Since the study of GW events is in the early stages, it is needless
to say that the multimessenger approach is exceedingly important in order to understand 
the nature of the production mechanisms.
Especially, mergers of NS-NS binaries are hypothesized
to be a possible origin of short gamma-ray bursts (sGRBs) (e.g. \citet{Pac86,Goo86,Eic89,Nar92,Moc93})
and thus the observation in the gamma-ray energy region is essential to understand the connection
between sGRBs to GW events.

We summarize the characteristics of six GW events during the first and second advanced LIGO-Virgo 
observing runs in Table \ref{tab:GWevents} with inferred parameters. 
We then
report the analysis of {\sl CALET}/CAL observations corresponding to
these gravitational events (except GW150914, which occurred before the start
of {\sl CALET} operations) in the gamma-ray energy region
as briefly summarized in Table \ref{tab:GWevents}. 

\begin{longrotatetable}
\begin{deluxetable}{ccccccccccl}

\tablecaption{Summary of CALET observations of gravitational events reported by the Virgo and LIGO scientific collaborations (BH: black hole, NS: neutron star) and representative results from CALET observation 
(see text for other time windows.)
\label{tab:GWevents} }
\tablehead{
\colhead{GW} & \colhead{Time} & \colhead{Location} & \colhead{Luminosity}
 & \colhead{Event} & \colhead{Ref.} & \multicolumn{4}{c}{CALET resullts [time window]} \\
  \cline{7-10}
 event & $T_0$ & area & distance  & Type & & Mode & Summed& \multicolumn{2}{c}{Upper limits (90\% C.L.)} \\
  \cline{9-10}
 & (UTC) & (deg$^2$) & (Mpc) & & & & LIGO & Energy flux & Luminosity \\
 &  &  & & & & & probability & (erg\,cm$^{-2}$s$^{-1}$) & (erg\,s$^{-1}$) 
}
\startdata
GW150914 & 2015-09-14 & 600 & $440^{+160}_{-180}$ & BH-BH & [a] & & & \multicolumn{2}{l}{Before operation}\\
 & 09:50:45 \\
GW151226 & 2015-12-26 & 850 & $440^{+180}_{-190}$ & BH-BH & [b] & LE & 15\% &  \multicolumn{2}{l}{[$T_0-525{\rm s},T_0+211{\rm s}$]}\\
 & 09:54:43  &&&&&&& $9.3\times10^{-8}$ & $2.3\times10^{48}$ \\
GW170104 & 2017-01-04 & 1200 & $880^{+450}_{-390}$ & BH-BH & [c] & HE & 30\% & \multicolumn{2}{l}{[$T_0-60{\rm s},T_0+60{\rm s}$]}\\
 & 10:11:58  &&&&&&&  $6.4\times10^{-6}$ & $6.2\times10^{50}$ \\
GW170608 & 2017-06-08 & 520 & $340^{+140}_{-140}$ & BH-BH  & [d] & HE & & \multicolumn{2}{l}{[$T_0-60{\rm s},T_0+60{\rm s}$]}\\
 & 02:01:16  &&&&&&&  \multicolumn{2}{l}{Out of FOV} \\
GW170814 & 2017-08-14 & 60 & $540^{+130}_{-210}$ & BH-BH & [e] & HE & & \multicolumn{2}{l}{[$T_0-60{\rm s},T_0+60{\rm s}$]}\\
 & 10:30:43  &&&&&&&  \multicolumn{2}{l}{Out of FOV} \\
GW170817  & 2017-08-17 & 28 & $40^{+8}_{-14}$ & NS-NS & [f] & HE & & \multicolumn{2}{l}{[$T_0-60{\rm s},T_0+60{\rm s}$]}\\
 & 12:41:04 &&&&&&& \multicolumn{2}{l}{Out of FOV} \\
\enddata

\medskip
Ref. [a] \citet{Abb16c}, [b] \citet{Abb16b}, [c] \citet{Abb17a}, [d] \citet{Abb17f}, [e] \citet{Abb17b}, [f] \citet{Abb17c}, 
\end{deluxetable}
\end{longrotatetable}

\section{Observation and Analysis}\label{sec:Analysis}

\subsection{CALET observation}

The CALorimetric Electron Telescope ({\sl CALET}) mission \citep{Tor15}
was launched and placed on the
Japanese Experiment Module-Exposed Facility of the International
Space Station (ISS) in 2015 August. At the LIGO trigger time
of GW150914, {\sl CALET} was in its commissioning phase and no
observational data were available. It was fully
{\bf functional} at the trigger times of GW151226, GW170104, GW170608, 
GW170814, and GW170817. 

\begin{figure}
\begin{center}
\includegraphics[width=.45\textwidth]{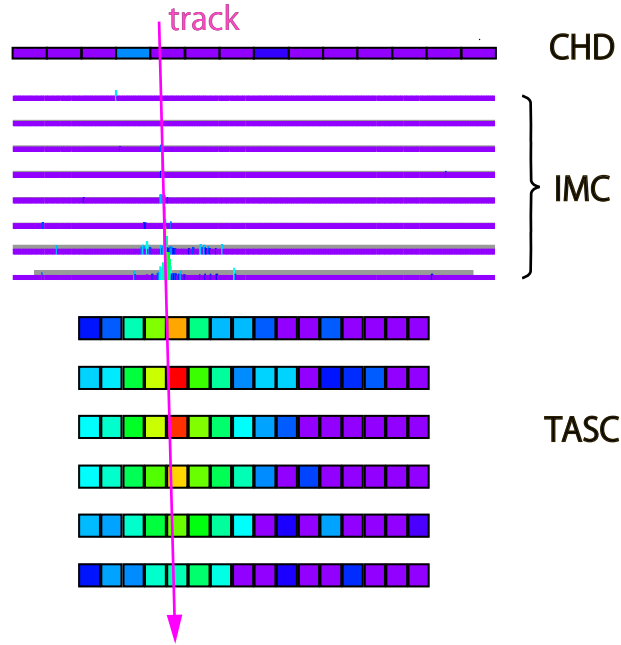}
\caption{Schematic cross-sectional view of {\sl CALET}/CAL with a sample event. 
We require that a track should cross the CHD (full area) 
and have at least a minimum path length in the TASC (see \citet{Can17,Can18} for details).}
\label{fig:CAL}
\end{center}
\end{figure}
There are two scientific instruments onboard {\sl CALET}: (1) The Calorimeter (CAL), the main
instrument, is a 30 radiation length deep calorimeter which can observe high-energy
lectrons in the energy range $\sim\!1$~GeV  -- $\sim\!20$~TeV, protons, helium, and
heavy nuclei in the energy range $\sim\!10$~GeV -- $\sim\!1000$~TeV and gamma-rays
in the energy range $\sim\!1$~GeV -- $\sim\!10$~TeV. The field of view (FOV) of CAL extends to
$\sim 45^\circ$ from the zenith direction. 
For gamma rays, the energy resolution and the angular resolution are
estimated as 3\% and $0.4^\circ$, respectively, at 10~GeV \citep{Mor13,Can17,Can18}.
CAL consists of three main components: the CHarge Detector (CHD), 
the IMaging Calorimeter (IMC), and the Total AbSorption Calorimeter (TASC)
(Figure \ref{fig:CAL}).
CHD is made up of a set of X and Y-direction arrays of 14
plastic scintillator paddles (32 mm $\times$ 10 mm $\times$ 450 mm);
IMC is composed of 8 layers of X- and Y-direction arrays of
448 scintillation fibers (SciFi, 1 mm $\times$ 1 mm $\times$ 448 mm)
separated by tungsten plates with a total thickness of 3 radiation lengths
($X_0$); and TASC is made of 6 layers of X- and Y-arrays
of 16 lead tungstate (PbWO$_4$ or PWO) scintillation crystals (19 mm $\times$ 20 mm
$\times$ 326 mm) with a total thickness of 27~$X_0$. (See \citet{Asa17} for details.)
The performance of CAL for gamma rays and initial results  
for steady gamma-ray sources are described in \citet{Can17,Can18}. 
(2) A companion instrument, the {\sl CALET}
Gamma-ray Burst Monitor (CGBM), monitors 
gamma-ray bursts (GRBs) using two different kinds of crystal
scintillators (LaBr$_3$(Ce) and Bi$_4$Ge$_3$O$_{12}$ (BGO)) to cover a wide energy
range (7~keV -- 20~MeV) \citep{Yam13}. Results from CGBM are presented separately \citep{Yam17}.

We use two trigger modes of CAL for gamma-ray analysis: a low-energy gamma-ray (LE-$\gamma$) mode 
with an energy threshold $\sim\!1$~GeV used at low geomagnetic latitudes 
and following a CGBM burst trigger, and a high-energy (HE) mode 
with a threshold $\sim10$~GeV used in normal operation for all particles
irrespective of geomagnetic latitude \citep{Asa18}. 
Around the trigger time of GW151226, between 03:30
and 03:43 UT, CAL was collecting regular scientific data 
under the LE-$\gamma$ mode.
The high voltages supplied to photomultipliers of CGBM detectors were set at
the nominal values around 03:22 UT and turned off around 03:43
UT to avoid a high background radiation area. No
CGBM on-board trigger was generated at the trigger time of GW151226.
For other GW events (GW170104, GW170608, GW170814 and GW170817), 
CAL was collecting data in the HE mode
since the ISS was in the high latitude region in its orbit. 
First results on the analysis of GW151226 have already been published
\citep{Adr16}, and here we describe results with a refined analysis.
We also give results on the comprehensive analysis of the CAL data for these four later events.

\subsection{Analysis of gamma-ray events in CALET/CAL}

The selection process of gamma-ray events used for the HE mode is essentially
the same as described in \citet{Mor13}. For the LE-$\gamma$ mode
the selection and analysis are fully described in
\citet{Can17,Can18}. Here we summarize the procedures briefly.

\paragraph{Offline trigger}
In order to remove the effects of variation in the hardware trigger
threshold and gains in the
flight data sample, energy deposit thresholds higher than those
nominally applied by the hardware trigger are imposed
off-line both for LE-$\gamma$ and HE modes.

\paragraph{Tracking}
Event tracks are reconstructed for the HE mode
using the EM track algorithm \citep{Aka13}
developed for the electron analysis which is a powerful method
for reconstructing electromagnetic showers.
For the LE-$\gamma$ mode we use
the CC track algorithm \citep{Can17,Can18} optimized for photons
with energies below 10 GeV. It begins by finding clusters of
hit fibers in the three bottom layers of IMC separately for
the X and Y-projections and extending the candidate tracks
to the upper layers of IMC. The trajectory with the highest 
total energy deposit is selected. 
In the HE mode, contained events passing through the CHD 
with track lengths in TASC in excess of 26.4~cm are subjected to 
further analysis; in LE-$\gamma$ mode, in order to maximize 
the FOV, we select well contained events whose tracks 
satisfy more sophisticated geometrical conditions \citep{Can17,Can18}.

\paragraph{Shower shape/hadronic rejection}
Low energy gamma-ray events can be mimicked by
albedo (i.e. upward moving) secondary charged pions
from hadronic interactions in the calorimeter or the support
structure. These events are vetoed by requiring that more energy be deposited 
in the bottom IMC layer than in the layer where pair conversion occurs.
Further rejection of events with showers
not consistent with a pure electromagnetic cascade is provided
by a cut on the IMC concentration, which uses the lateral
spread of the energy deposit distribution in the lower layers
of IMC. 

In order to reject hadronic events we utilize the $K$ parameter defined as
$$
K = \log_{10}F_E + R_E/2\,{\rm cm}
$$
where $F_E$ is the fractional energy deposit in the bottom TASC
layer with respect to the total energy deposit sum in the TASC
and $R_E$ is the second moment of the lateral energy deposit
distribution in the top layer of TASC.
This method is developed for the derivation of the electron flux
and is designed to exploit the larger spread and slower development
of proton showers due to penetrating secondary pions \citep{Adr17}.

\paragraph{Zero charge identification}
In order to select events consistent with zero primary charge,
cuts are made on the energy deposits in CHD and upper IMC layers.
These requirements are designed to veto charged particle events
effectively. We require one of three filters utilizing CHD and upper IMC layers
(see \citet{Can17,Can18} for detail).

\medskip
In addition, as described in detail in \citet{Can17,Can18}, we have to reject
gamma-ray candidate events which are generated in the ISS structures 
such as the Japanese Experiment Module)
to remove events generated in interactions of cosmic rays 
with these structures, which create gamma-ray event clusters
clearly seen in our FOV.
After these selections, incident gamma-ray energies are derived 
from the deposited energies based on Monte Carlo simulations, 
pre-flight accelerator calibrations, and in-flight non-interacting 
penetrating particle events.

\begin{figure}
\begin{center}
\includegraphics[width=.65\textwidth]{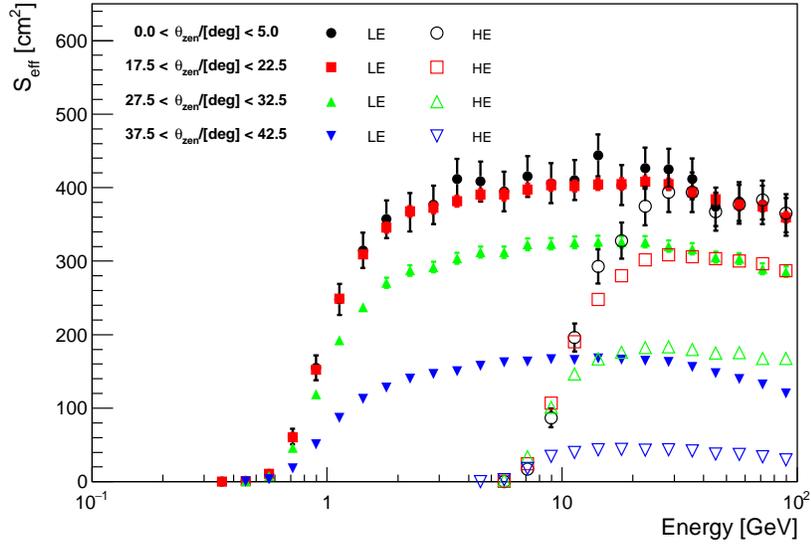}
\caption{Effective area of {\sl CALET}/CAL as a function of gamma-ray energy in the 
low-energy gamma-ray mode (LE-$\gamma$) and 
high-energy mode (HE). Four ranges of incident zenith angles ($\theta_{\rm zen}$) are assumed. 
Statistical uncertainties due to Monte Carlo statistics are shown by error bars.
\label{fig:effarea}
}
\end{center}
\end{figure}

Based on the CALET simulation studies \citep{Mor13, Can17,Can18}, 
the gamma-ray efficiency reaches its maximum
around 10 GeV with an efficiency of 48\% relative to an area of 
the TASC top layer (excluding a 1.9~cm margin around the outside) 
for normal incidence, after applying the event selections described above.
This figure is to be compared with a pair creation probability 
of 54\% in 1 radiation length, which is approximately the thickness required 
to be tracked in at least 3 layers in IMC, and implies
a high efficiency in the gamma-ray event reconstruction and selection processes.
The effective areas for four ranges of incident angles are shown 
in Figure \ref{fig:effarea} as a function of gamma-ray energy.

The GeV sky is rather bright along the Galactic plane due to the
Galactic diffuse gamma-ray radiation with Galactic and extragalactic individual sources, 
and there is a residual all-sky emission component called the isotropic diffuse
gamma-ray background. These gamma-rays are a source of background
in a search for gamma-ray emission associated with GW events.
The expected number of background events
in the time window used in our search was calculated
using a prediction based on the Fermi LAT Pass 8 measurements%
\footnote{We utilized a gamma-ray skymap in the
energy range 1 -- 100 GeV created using the archival data 
for the dates 2008-08-04 through 2017-03-12
available via \url{https://fermi.gsfc.nasa.gov/ssc/data/access/lat/}.
}. 
As shown by \citet{Can17,Can18}, the CALET measurement is in reasonable agreement with the LAT result.

The upper limit of the CAL observation in the time windows 
is estimated as follows: First, we calculate the effective
area as a function of gamma-ray energy, and the resultant energy-dependent 
exposure map in the time window for the
corresponding energy region depending on the trigger mode (LE-$\gamma$ or HE). 
In the case of a null event, we estimate the upper limit on the gamma-ray flux corresponding
to 2.44 events (the 90\% confidence limit for a null
observation) assuming a power-law spectrum with a single photon
index of $-2$ by using the calculated exposure map.
The photon index, $-2$, is taken as a typical value for 
{\sl Fermi}-LAT GRBs in the GeV energy range \citep{Ack13}.

\subsection{GW151226\,%
\footnote{The result shown here is an improved version based
on more refined analysis compared with that presented in our previous paper \citep{Adr16}.}}
\label{sec:GW151226}

We searched for gamma-ray events associated with GW151226
using the CAL data in the time window
$[T_0-525\,{\rm s},T_0 +211\,{\rm s}]$ around the LIGO trigger time ($T_0$), the time period when the CAL was
operational in the LE-$\gamma$ mode with an
energy threshold of 1 GeV.
We analyzed the full length of this window in order to  
perform the most sensitive search possible with increased statistics.

Expected number of contaminated background gamma-rays is small 
because the searched area of the sky for the GW151226 counterpart is significantly 
apart from the Galactic plane. 
In fact, the number of expected background evens is 0.051 in this time window
for the sky region covering 25\% of the summed LIGO probabilities;
i.e., the CAL observation is almost background-free for such a short time period. 
No candidates were found in this time window and sky region, resulting in an upper limit is calculated
as described in the previous section.

\begin{figure}
   \begin{center}
    \includegraphics[width=14cm]{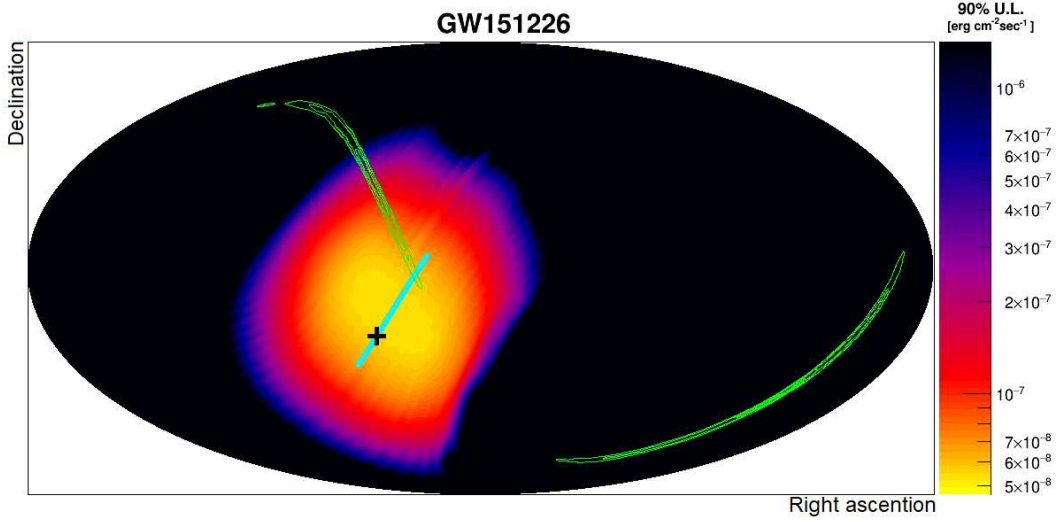}
   \end{center}
 \caption{90\% C.L.\ upper limit on GW151226 energy flux in the energy region 1 -- 10~GeV 
and time window [$T_0-525$~s, $T_0+211$~s] shown in the equatorial coordinates. 
Thick cyan line shows the locus of the FOV center of CAL, and the plus symbol is that at $T_0$.  
Also shown by green contours is the localization significance map of the GW151226 signal reported by LIGO.}
 \label{fig:upperlimit151226}
\end{figure}

Figure \ref{fig:upperlimit151226} shows the sky map of the 90\%-confidence-level upper
limit on the gamma-ray flux. The estimated upper
limit is $9.3\times10^{-8}$~erg\,cm$^{-2}$\, s$^{-1}$ (90\% C.L.)
in the 1 -- 10 GeV region where the coverage of CAL reaches 15\% of the 
integrated LIGO probability ($\sim1.1$~sr).
If we enlarge the sky region to contain 25\% of the LIGO integrated probability,
the upper limit is $2.8\times10^{-7}$~erg\,cm$^{-2}$\, s$^{-1}$
in the same energy region.
The luminosity upper limit set by CAL is estimated as
$2.3\,(6.8)\times10^{48}$~erg\,s$^{-1}$ assuming a luminosity distance of 440 Mpc
for coverage of $\sim15\,(25)$\% of the LIGO integrated probability regions. 
By comparison, the upper limit in the energy flux in the 0.1 -- 1 GeV region 
as reported by {\sl Fermi}-LAT (assuming a power-law spectrum 
with a single photon index of $-2$) is $3\times 10^{-10}$~erg\,cm$^{-2}$\,s$^{-1}$ 
(95\% C.L.) for the time window $[T_0, T_0 + 1\times 10^4 \,{\rm s}]$ \citep{Rac16}, 
corresponding to $\sim 4\times10^{-9}$~erg\,cm$^{-2}$\,s$^{-1}$
for the 736-s time window of the CAL in the LE$\gamma$ mode for this GW event.

We also calculate upper limits on energy flux of gamma rays in smaller time windows
since we do not know the time profile of the possible electromagnetic emission
which accompanies gravitational wave events.
When we set the window as $[T_0-60\,{\rm s},T_0+60\,{\rm s}]$,
the upper limit in the 1 -- 10 GeV region is $9.4\,(20)\times10^{-7}$~erg\,cm$^{-2}$s\,$^{-1}$
for the integrated LIGO probabilities inside the CAL FOV of $\sim15\,(25)$\%.
If we set the window as $[T_0-1\,{\rm s},T_0+1\,{\rm s}]$,
the upper limit in the 1 -- 10 GeV region is $5.3\times10^{-5}$~erg\,cm$^{-2}$\,s$^{-1}$
for the LIGO integrated probabilities in the CAL FOV at the level of $\sim15$\%.

\subsection{GW170104}

\begin{figure}
   \begin{center}
       \includegraphics[width=14cm]{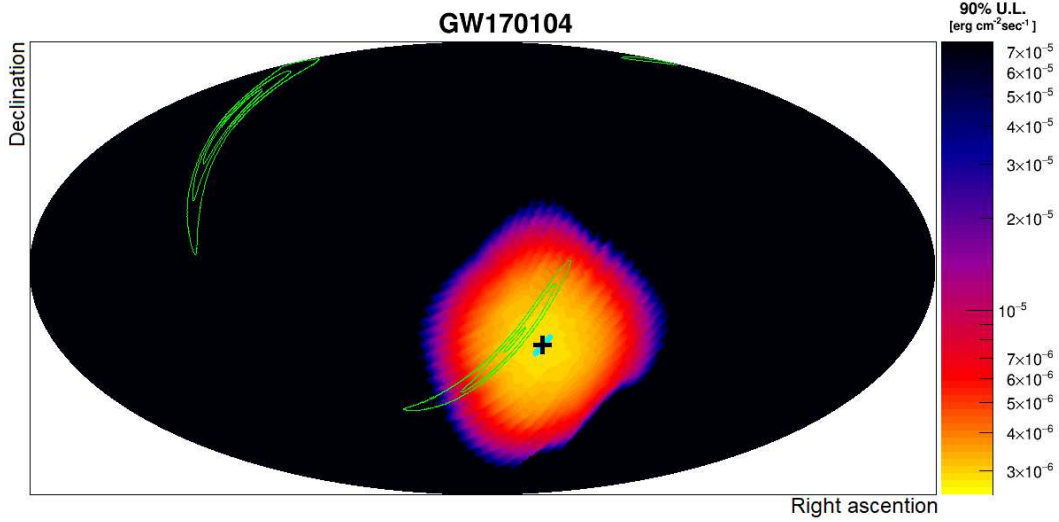} \end{center}
 \caption{90\% C.L.\ upper limit on GW170104 energy flux in the energy region 10 -- 100~GeV 
and time window [$T_0-60$~s, $T_0+60$~s] shown in the equatorial coordinates. 
Thick cyan line shows the locus of the FOV center of CAL, and the plus symbol is that at $T_0$.  
Also shown by green contours is the localization significance map of the GW170104 signal reported by LIGO.}
 \label{fig:upperlimit170104}
\end{figure}

For the time period around the trigger time ($T_0$) corresponding to GW170104, 
CAL was running in the HE mode with an energy threshold of 10~GeV.
Gamma-ray events have been searched for
using the CAL data in the time window $[T_0-60\,{\rm s},T_0+60\,{\rm s}]$ but no candidates were found.
The estimated number of background events expected in this time window is $7.8\times10^{-4}$.
We calculated an upper limit on the gamma-ray energy flux
of $6.4\times10^{-6}$~erg\,cm$^{-2}$\,s$^{-1}$ at 90\% C.L. in the 10 -- 100~GeV energy region
for the sky region covering $30$\% of the integrated LIGO probabilities (Figure \ref{fig:upperlimit170104}).
This upper limit corresponds to $6.2\times10^{50}$~erg\,s$^{-1}$ assuming a luminosity distance of 880~Mpc.
If we set a narrower time window as [$T_0-1$, $T_0+1$~s], 
the estimated number of background events is $1.2\times10^{-5}$
and the upper limit is
$4.3\times10^{-4}$~erg\,cm$^{-2}$\,s$^{-1}$ for the flux and $4.1\times10^{52}$~erg\,s$^{^-1}$ 
for the luminosity (90\% C.L.) assuming the same sky region.

We note that {\sl AGILE} reported a weak ($4.4\sigma$) event
lasting about 32~ms and occurring $0.46\pm0.05$~s before $T_0$
in the omni-directional MCAL data in the 0.4 -- 100~MeV region \citep{Ver17a},
while other searches for high-energy emission yielded upper limits only.

\subsection{GW170608}

\begin{figure}
   \begin{center}
    \includegraphics[width=14cm]{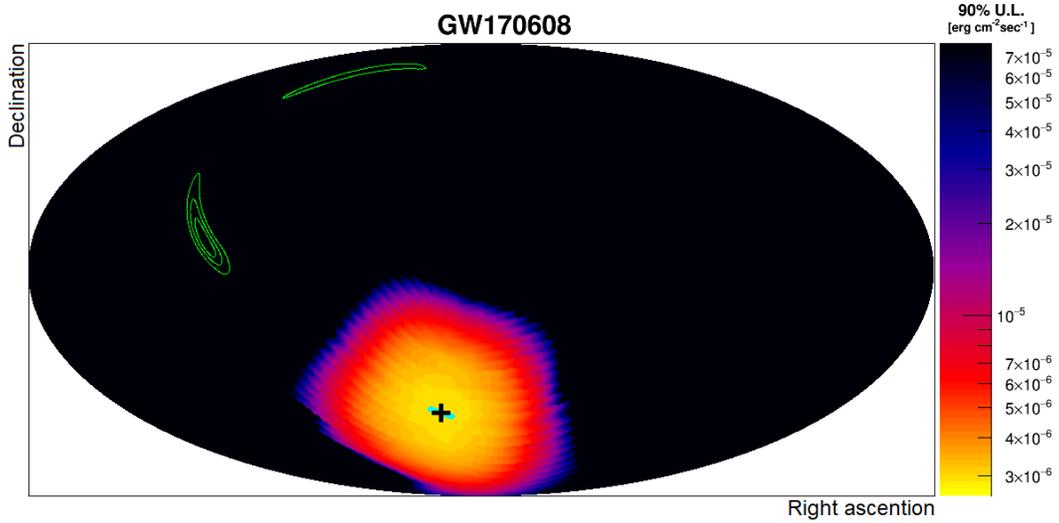}
   \end{center}
 \caption{90\% C.L.\ upper limit on GW170608 energy flux in the energy region 10 -- 100~GeV 
and time window [$T_0-60$~s, $T_0+60$~s] shown in the equatorial coordinates. 
Thick cyan line shows the locus of the FOV center of CAL, and the plus symbol is that at $T_0$.  
Also shown by green contours is the localization significance map of the GW170608 signal reported by LIGO.}
 \label{fig:upperlimit170608}
\end{figure}

For the time period around the trigger time ($T_0$) corresponding to GW170608, 
CAL was running in the HE mode with an energy threshold of 10~GeV.
Gamma-ray events have been searched for
using the CAL data in the time window $[T_0-60\,{\rm s},T_0+60\,{\rm s}]$ but no candidates were found.
Unfortunately, the sky coverage of CAL did not include the region of
the localization (520 deg$^2$) determined with two interferometric detectors
as shown in Figure \ref{fig:upperlimit170608}. 

We note that {\sl Fermi}-LAT reported a weak ($3.5\sigma$) excess around the LIGO location
area in the $[T_0,T_0+1~{\rm ks}]$ window in the energy region above 100 MeV \citep{Omo17}, but others
reported only upper limits for high-energy emission for this GW event.

\subsection{GW170814}

\begin{figure}
   \begin{center}
    \includegraphics[width=14cm]{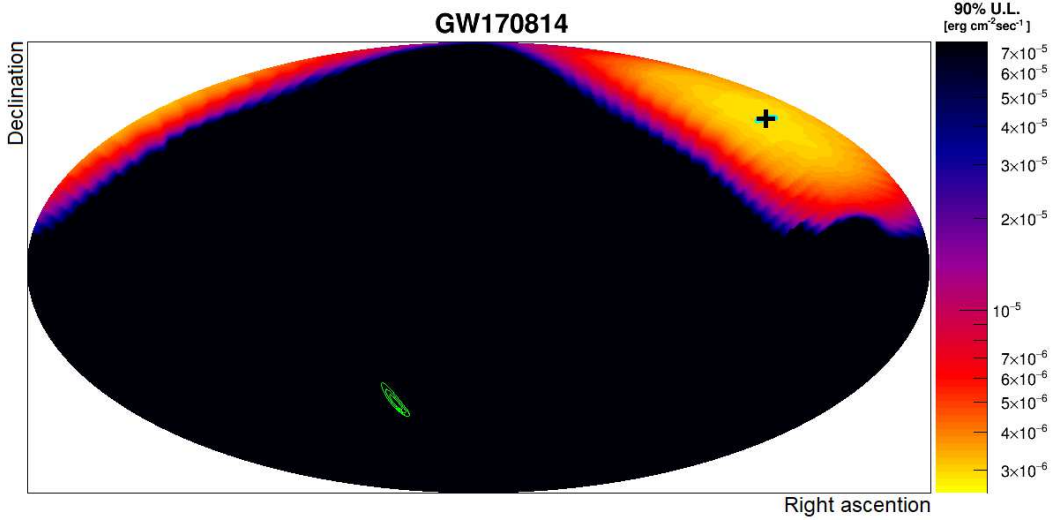}
   \end{center}
 \caption{90\% C.L.\ upper limit on GW170814 energy flux in the energy region 10 -- 100~GeV 
and time window [$T_0-60$~s, $T_0+60$~s] shown in the equatorial coordinates. 
Thick cyan line shows the locus of the FOV center of CAL, and the plus symbol is that at $T_0$.  
Also shown by green contours is the localization significance map of the GW170814 signal reported by LIGO/Virgo.}
 \label{fig:upperlimit170814}
\end{figure}

For the time period around the trigger time ($T_0$) corresponding to GW170814, 
CAL was running in the HE mode with an energy threshold of 10~GeV.
Gamma-ray events have been searched for
using CAL data in the time window $[T_0-60\,{\rm s},T_0+60\,{\rm s}]$ but no candidates were found.
Unfortunately, the sky coverage of CAL did not include the rather small region of
the localization (60 deg$^2$) determined with three interferometric detectors,
as shown in Figure \ref{fig:upperlimit170814}. 

We note that {\sl INTEGRAL}/SPI-ACS reported a weak $3.5\sigma$ excess in the
$[T_0-1.5\,{\rm s},T_0+8.5\,{\rm s}]$ window \citep{Poz17}, but this was not confirmed 
by an independent analysis \citep{Sav17}. Other reports gave only upper limits
on high-energy emission for this GW event.

\subsection{GW170817}

\begin{figure}
   \begin{center}
    \includegraphics[width=14cm]{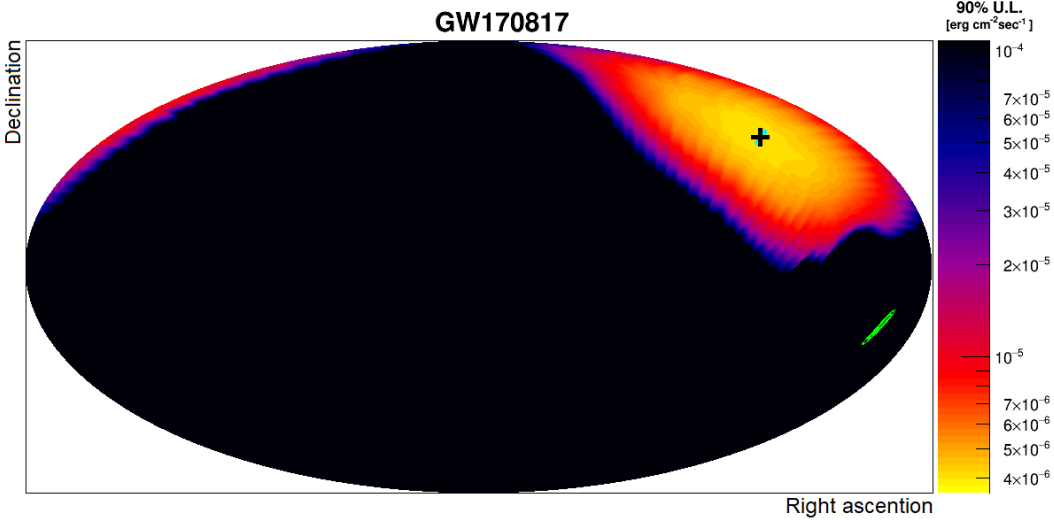}
   \end{center}
 \caption{90\% C.L.\ upper limit on GW170817 energy flux in the energy region 10 -- 100~GeV 
and time window [$T_0-60$~s, $T_0+60$~s] shown in the equatorial coordinates. 
Thick cyan line shows the locus of the FOV center of CAL, and the plus symbol is that at $T_0$.  
Also shown by green contours is the localization significance map of the GW170817 signal reported by LIGO/Virgo.}
 \label{fig:upperlimit170817}
\end{figure}

1.7 s after the trigger due to the LIGO-Virgo event GW170817 ($T_0$), 
{\sl Fermi}-GBM and {\sl INTEGRAL} detected GRB 170817A with $T_{90}$ duration 
$2.0 \pm 0.5$~s \citep{Abb17d}. 
For the time period around GW 170817, 
CAL was running in the HE mode with an energy threshold of 10~GeV.
Gamma-ray events have been searched for
using the CAL data in the time window $[T_0-60\,{\rm s},T_0+60\,{\rm s}]$ but no candidates were found.
Unfortunately, the sky coverage of CAL did not include the rather small region of
the localization (28 deg$^2$) determined with three interferometric detectors,
as shown in Figure \ref{fig:upperlimit170817}. It is reported the gravitational wave signal
started about 100~s before $T_0$, but it was also out of the field-of-view of CAL during this period.

We have also searched for possible delayed signal from this merger event \citep{Mur17a}.
In the two-month period (Aug. 17 -- Oct. 16, 2017) after the event
we had no gamma-ray candidate around the
direction of its counterpart object (NGC 4993) \citep{Abb17e}, and obtained 90\% C.L. upper limits on the
energy flux of $1.2\times10^{-10}$~erg\,cm$^{-2}$s$^{-1}$ 
($4.0\times10^{-10}$~erg\,cm$^{-2}$s$^{-1}$) for gamma rays above 1 GeV (10 GeV)
using the LE-$\gamma$ mode (the HE mode).
This upper limit corresponds to $2.4\times10^{43}$~erg\,s$^{-1}$
($8.0\times10^{43}$~erg\,s$^{-1}$) assuming a luminosity distance of 40~Mpc.

\section{Future Prospects}

Identifying the electromagnetic counterpart of a gravitational-wave event would be 
a key discovery to constrain the origin of the event. 
The detection of multiwavelength radiation in association with GW170817 \citep{Abb17e}
was a huge step to open a new window of astronomy.
In particular the detection of a gamma-ray burst, GRB170817A, observed $\sim1.7$~s 
after GW170817 by {\sl Fermi}-GBM and {\sl INTEGRAL} \citep{Abb17d}, provides new insight into the origin
of short gamma-ray bursts. The association of GW170817 and GRB170817A can be
interpreted as a merger of a neutron star-neutron star binary, which is hypothesized
to be a possible origin of short gamma-ray bursts as discussed in section 1.
However, GRB170817A, which is the closest short GRB ever observed, is 2 to 6 orders of magnitude
less energetic than other bursts with known distances.

The underluminous nature of GRB 170817A
may imply that the gamma-rays detected with {\sl Fermi}-GBM are off-axis emission 
from a typical short GRB \citep{Abb17c, Ale18, Iok17, Laz17, Mar17, Tro17}.
Although the (50 -- 300 keV)/(10 -- 50 keV) hardness ratio 
 is small (``soft''), the spectral peak energy is close to the lower end 
of the typical value in spite of the off-axis observation (but see \citet{Kis17} for example).
The rising X-ray and radio afterglow lightcurves as far as $\sim 100$ days \citep{Moo17, Rua17} 
are also difficult to explain with an off-axis afterglow model with a simple top-hat jet 
(or they may suggest a structured jet).
GRB 170817A may belong to another population of gamma-ray transient phenomena 
other than the short GRB as proposed by \citet{Bro17}, \citet{Got17}, \citet{Kas17}, 
\citet{Mur17}, and \citet{Asano18}.
In such cases, the expected gamma-ray flux in the GeV range is not constrained by 
the previous short GRB observations.

The fluence of GRB 170817A in the keV-MeV energy band
was observed to be $(1.4\pm0.3)\times10^{-7}$~erg\,cm$^{-2}$ \citep{Abb17c}. 
Had the same level of fluence been present in the GeV energy band, it could have be detected
by GeV gamma-ray detectors in operation at that time. 
However, {\sl Fermi}-LAT was entering the South Atlantic Anomaly
and was not collecting data until about $10^3$ seconds after GRB 170817A \citep{Fer17},
and {\sl AGILE} started observation after about $10^3$ seconds \citep{Ver17b}.
It was also out of the field-of-view of {\sl CALET} as reported above. Thus unfortunately there
is no limit in the GeV band around the trigger time of GW170817.

Regarding sGRB events in general, some events have been observed 
to emit high-energy ($> 100$~MeV) gamma-rays (e.g., GRB 081024B, 
\citet{Abd10}; and GRB 090510, \citet{Ack10}). 
Their fluence in the high-energy band could be comparable to that in the hard X-ray band \citep{Abd10}.
However, the fraction of GRBs showing high-energy emission observed to date is fairly low \citep{Ack13}.
This could be due to intrinsic properties associated with the gamma-ray emission mechanism,
but another reason could be the limited field-of-view of GeV gamma-ray
detectors ($2\sim3$ sr). 

\begin{figure}
   \begin{center}
    \includegraphics[width=14cm]{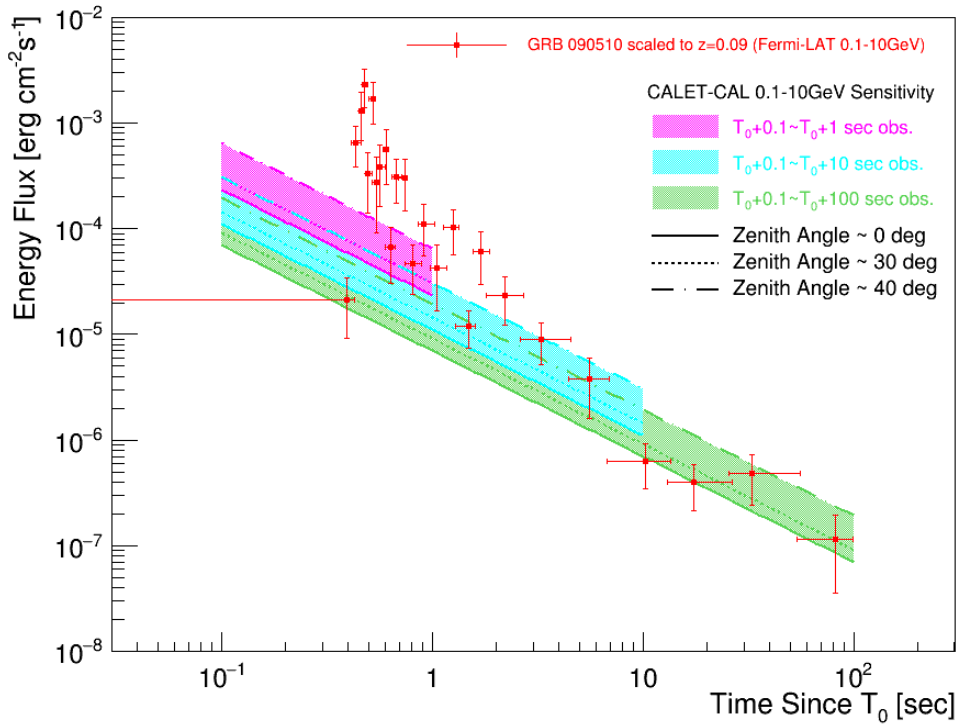}
   \end{center}
 \caption{{\sl CALET}/CAL sensitivity to obtain 1 event for a transient source assuming the energy spectrum 
proportional to $E^{-2}t^{-1}$, where $E$ is the energy and $t$ is the time after $T_0$,
in the energy region 0.1 -- 10~GeV.
Despite the lack of sensitivity to sub-GeV gamma rays in the CAL, 
the 0.1 -- 1 GeV band is included in this calculation of the limit to compare to the {\sl Fermi}-LAT light
curve since the energy flux is sensitive to the range over which it is integrated.
Shaded areas show energy-flux sensitivities assuming observations of 1, 10 and 100 s duration
for a source around the zenith, and dotted and dot-dashed lines show those for a 
source around 30$^\circ$ and 40$^\circ$ from zenith, respectively. Also shown by points
are the observed light curve of GRB 090510 by {\sl Fermi}-LAT ,
which is a short-hard GRB with an additional hard power-law component from 10~keV to GeV energies
\citep{Fer17}, scaled to $z=0.09$, 
the nominal redshift of the first LIGO event GW150914 as calculated by \citet{Ack16}.  }
 \label{fig:sensitivity}
\end{figure}

Figure \ref{fig:sensitivity} shows the sensitivity of {\sl CALET}/CAL to obtain 1 event 
assuming an observation of 1, 10 and 100 s duration.
The typical energy range of the on-axis gamma-ray emission from NS-NS mergers 
can be higher than that of the short GRB emission from GRB170817A.
Although the effective area of {\sl CALET}/CAL is smaller than 
that of {\sl Fermi}-LAT, the fields-of-view of the two detectors are comparable.
As the sensitivity of laser interferometers is expected to increase in coming years,
the number of gamma-ray transients associated with GW events falling into 
the possible new population mentioned above will also increase. 
If their spectra extend to GeV energies, they could be easily detectable
as shown in Figure \ref{fig:sensitivity}.
Thus {\sl CALET}/CAL could contribute to constrain the GeV emission 
from a nearby NS-NS merger simultaneously with a GW signal in the near future.
Monitoring the GeV sky with CALET, with its mission scheduled to continue 
for three more years, may complement the coverage by other missions 
and may help to study unexplored high-energy emission from future transient events.

\acknowledgments

We would like to thank the anonymous referee for comments and suggestions that materially improved the paper.
We gratefully acknowledge JAXA's contributions for {\sl CALET} development and operation on ISS. 
We express our sincere thanks to ASI and NASA for their support to the {\sl CALET} project. 
This work is partially supported by JSPS Grant-in-Aid for Scientific Research (S) Number 26220708, 
JSPS Grant-in-Aid for Scientific Research (B) Number 17H02901, 
JSPS Grant-in-Aid for Scientific Research (C) Number 16K05382 
and MEXT-Supported Program for the Strategic Research
Foundation at Private Universities (2011-2015) S1101021 in Waseda University. 
This work is also supported in part
by MEXT Grant-in-Aid for Scientific Research on Innovative Areas Number 24103002. 
US {\sl CALET} work is supported
by NASA under RTOP 14-APRA14-0075 (GSFC) and grants NNX16AC02G (WUStL), NNX16AB99G (LSU), and
NNX11AE06G (Denver).


\begin{thebibliography}{}
\bibitem[Abbott et al.(2016a)]{Abb16a}
Abbott, B.~P. et al. 2016a, \prl, 116, 061102
\bibitem[Abbott et al.(2016b)]{Abb16b}
Abbott, B.~P. et al. 2016b, \prl, 116, 241103
\bibitem[Abbott et al.(2016c)]{Abb16c}
Abbott, B.~P. et al. 2016c, Phys. Rev. X, 6, 041015
\bibitem[Abbott et al.(2017a)]{Abb17a}
Abbott, B.~P. et al. 2017a, \prl ,118, 221101
\bibitem[Abbott et al.(2017b)]{Abb17b}
Abbott, B.~P. et al. 2017b, \prl, 119, 141101
\bibitem[Abbott et al.(2017c)]{Abb17c}
Abbott, B.~P. et al. 2017c, \prl, 119, 161101
\bibitem[Abbott et al.(2017d)]{Abb17d}
Abbott, B.~P. et al. 2017d, \apjl, 848, L13
\bibitem[Abbott et al.(2017e)]{Abb17e}
Abbott, B.~P. et al. 2017e, \apjl, 848, L12
\bibitem[Abbott et al.(2017f)]{Abb17f}
Abbott, B.~P. et al. 2017f, \apjl, 851, L35 
\bibitem[Abdo et al.(2010)]{Abd10}
Abdo, A.~A. et al. 2010, \apj, 712, 558
\bibitem[Ackermann et al.(2010)]{Ack10}
Ackermann, M. et al. 2010, \apj, 716, 1178 
\bibitem[Ackermann et al.(2013)]{Ack13}
Ackermann, M. et al. 2013, \apjs, 209, 11 
\bibitem[Ackermann et al.(2016)]{Ack16}
Ackermann, M. et al. 2016, \apjl, 823, L2  
\bibitem[Acero et al.(2016)]{Ace16}
Acero, F. et al. 2016, \apjs, 223, 26
\bibitem[Adriani et al.(2016)]{Adr16}
Adriani, O. et al. 2016, \apjl, 829, L20
\bibitem[Adriani et al.(2017)]{Adr17}
Adriani, O. et al. 2017, \prl, 119, 181101
\bibitem[Akaike et al.(2013)]{Aka13}
Akaike, Y. for the CALET Collaboration 2013, in Proc. 33rd ICRC (Rio de Janeiro, Brazil, 2013), 0726 (pp. 2162--2165).
\bibitem[Alexander et al.(2018)]{Ale18}
Alexander, K. D., Berger, E., Fong, W., et al. 2017, \apjl, 848, L21 
\bibitem[Asano \& To(2018)]{Asano18}
Asano, K., \& To, S. 2018, \apj, 852, 105 
\bibitem[Asaoka et al.(2017)]{Asa17}
Asaoka, Y. et al. 2017, Astropart. Phys. 91, 1
\bibitem[Asaoka et al.(2018)]{Asa18}
Asaoka, Y. , Ozawa, S., Torii, S. et al. 2018, Astropart. Phys. 100, 29
\bibitem[Bromberg et al.(2017)]{Bro17}
Bromberg, O., Tchekhovskoy, A., Gottlieb, O., Nakar, E., \& Piran, T. 2017, arXiv:1710.05897
\bibitem[Cannady et al.(2017)]{Can17}
Cannady, N. for the CALET Collaboration 2017, in Proc. 35th ICRC (Busan, Korea, 2017) (PoS (ICRC2017) 720) 
\bibitem[Cannady et al.(2018)]{Can18}
Cannady, N. for the CALET Collaboration 2018, submitted for publication
\bibitem[De Mink and King(2017)]{deM17}
de Mink, S.~E. and King, A. 2017, \apjl, 839, L7
\bibitem[Eichler et al(1989)]{Eic89}
Eichler, D., Livio, M., Piran, T. \& Schramm, D. N. 1989, \nat, 340, 126
\bibitem[Einstein(1916,1918)]{Ein16}
Einstein, A., 1916, 1918, Sitzungsberichte der K\"{o}niglich Preussischen Akademie 
der Wissenschaften Berlin, 1016, 688--696; {\it ibid.} 1918, 154--167.
\bibitem[Fermi-LAT collaboration(2017)]{Fer17}
{\sl Fermi}-LAT collaboration 2017, arXiv:1710.05450
\bibitem[Fern\'{a}ndez and Metzger(2016)]{Fer16}
Fern\'{a}ndez, R. \& Metzger, B.D. 2016, Ann. Rev. Nucl. Part. Sci., 66, 23
\bibitem[Goodman(1986)]{Goo86}
Goodman, J. 1986, \apj, 308, L47
\bibitem[Gottlieb et al.(2017)]{Got17}
Gottlieb, O., Nakar, E., Piran, T., \& Hotokezaka, K. 2017, arXiv:1710.05896
\bibitem[Ioka \& Nakamura(2018)]{Iok17}
Ioka, K., \& Nakamura, T. 2018, Prog.\ Theor.\ Exp.\ Phys., 043E02
\bibitem[Kasliwal al.(2017)]{Kas17}
Kasliwal, M. M., Nakar, E., Singer, L. P., et al. 2017, Science, 358, 1559
\bibitem[Kisaka et al.(2017)]{Kis17}
Kisaka, S., Ioka, K., Kashiyama, K. \& Nakamura, T. 2017, arXiv:1711.00243
\bibitem[Lazzati et al.(2018)]{Laz17}
Lazzati, D., Perna, R., Morsony, B. J., et al. 2018, \prl, 120, 241103
\bibitem[Margutti et al.(2017)]{Mar17}
Margutti, R., Berger, E., Fong, W., et al. 2017, \apjl, 848, L20
\bibitem[Mochkovitch et al.(1993)]{Moc93}
Mochkovitch, R., Hernanz, J., Isern, J. \& Martin, X. 1993, \nat, 361, 236
\bibitem[Mooley et al.(2017)]{Moo17}
Mooley, K. P., Nakar, E., Hotokezaka, K., et al. 2017, \nat, 554, 207
\bibitem[Mori et al.(2013)]{Mor13}
Mori, M. for the CALET Collaboration 2013, in Proc. 33rd ICRC (Rio de Janeiro, Brazil, 2013), 0248 (pp. 1185--1188).
\bibitem[Murase et al.(2018)]{Mur17a}
Murase, K. et al. 2018, \apj, 854, 60
\bibitem[Murguia-Berthier et al.(2017)]{Mur17}
Murguia-Berthier, A., Ramirez-Ruiz, E., Kilpatrick, C. D., et al. 2017, \apjl, 848, L34
\bibitem[Narayan et al.(1992)]{Nar92}
Narayan, R. Paczynsky, B. \& Piran, T. 1992, \apjl, 395, L83 
\bibitem[Omodei et al.(2017)]{Omo17}
Omodei, N. et al. 2017, GCN, 21227
\bibitem[Pacynski(1986)]{Pac86}
Pacy\'{n}ski, B. 1986, \apj, 308, L43
\bibitem[Phinney(2009)]{Phi09}
Phinney, E.~S. 2009, in New Worlds, New Horizons in Astronomy and Astrophysics
(The National Academies Press) / arXiv:0903.0098 
\bibitem[Pozanenko et al.(2017)]{Poz17}
Pozanenko, A. et al. 2017, GCN, 21476
\bibitem[Racusin et al.(2017)]{Rac16}
Racusin, J.L. et al. 2017, \apj, 835, 82
\bibitem[Rosswog(2015)]{Ros15}
Rosswog, S. 2015, Int. J. Mod. Phys., 24, 1530012
\bibitem[Ruan et al.(2017)]{Rua17}
Ruan, J. J., Nynka, M., Haggard, D., Kalogera, V., \& Evans, P. 2017, arXiv:1712.02809
\bibitem[Savchenko et al.(2017)]{Sav17}
Savchenko, V. et al. 2017, GCN, 21478
\bibitem[Torii et al.(2015)]{Tor15}
Torii, S. for the CALET Collaboration 2015, in Proc. 34th ICRC (Hague, Netherland, 2015) (PoS (ICRC2015) 581)
\bibitem[Troja et al.(2017)]{Tro17}
Troja, E., Piro, L., van Eerten, H., et al. 2017, \nat, 551, 71
\bibitem[Veracchia et al.(2017a)]{Ver17a}
Verrecchia, F. et al. 2017a, \apjl, 847, L20
\bibitem[Verrecchia et al.(2017b)]{Ver17b}
Verrecchia, F. et al. 2017b, \apjl, 850, L27
\bibitem[Yamaoka et al.(2013)]{Yam13}
Yamaoka, K. for the CALET Collaboration 2013, in Proc. 7th Huntsville Gamma-Ray 
Burst Symposium (Nashville, USA, 2013), paper 41/eConf Proceedings C1304143
\bibitem[Yamaoka et al.(2017)]{Yam17}
Yamaoka, K. for the CALET Collaboration 2017, in Proc. 35th ICRC (Busan, Korea, 2017) (PoS (ICRC2017) 614)

\end{thebibliography}
\end{document}